\documentclass[12pt]{iopart}

\usepackage{relsize}
\usepackage{iopams}
\usepackage{graphicx}

\begin{document} 

\title[]{Conditions for the Invar effect in Fe$_{1-x}A_{x}$ ($A={\rm Pt},{\rm Ni}$)}

\author{Fran\c{c}ois Liot$^{1,2,3}$\footnote{Present address: Norinvar, 59 la rue, 50110 Bretteville, France.}}
\address{$^{1}$ Norinvar, 59 la rue, 50110 Bretteville, France}
\address{$^{2}$ Department of Computational Materials Design (CM), Max-Planck-Institut f{\"u}r Eisenforschung GmbH, D-40237 D{\"u}sseldorf, Germany}
\address{$^{3}$ Department of Physics, Chemistry, and Biology (IFM), Link{\"o}ping University, SE-581 83 Link{\"o}ping, Sweden}
\ead{f.liot@norinvar.com}

\begin{abstract}
We present a necessary condition under which a collinear ferromagnet Fe$_{1-x}A_{x}$ ($A={\rm Pt},{\rm Ni}$) with disordered face-centered-cubic structure exhibits the Invar effect. The condition involves the rate at which the fraction of Fe moments that are antiferromagnetically aligned with the magnetization fluctuates as the system is heated, $dx^{{\rm Fe}\downarrow}/dT$. Another contributing factor is the magnetostructural coupling $\kappa=-(1/V)(\partial V / \partial x^{{\rm Fe}\downarrow})_{T}$, where the volume $V(T, x^{{\rm Fe}\downarrow})$ corresponds to a homogeneous ferromagnetic state, a partially disordered local moment state, or a disordered local moment state depending on the value of $x^{{\rm Fe}\downarrow}$. According to the criterion, the Invar phenomenon occurs only when the thermal expansion arising from the temperature dependence of the fraction of Fe moments which point down $-1/3\,\kappa\,dx^{{\rm Fe}\downarrow}/dT$ compensates for the thermal expansion associated with the anharmonicity of lattice vibrations in a wide temperature interval. Upon further investigation, we provide evidence that only alloys with strong magnetostructural coupling at zero Kelvin can show the Invar effect.

\end{abstract}


\maketitle

\section{Introduction} \label{sec_intro}

Disordered face-centered-cubic (fcc) Fe$_{0.72}$Pt$_{0.28}$ and Fe$_{0.65}$Ni$_{0.35}$ alloys have remained at the forefront of condensed matter theory for more than sixty years, owing to their rich variety of intriguing physical properties. Their linear thermal expansion coefficient (LTEC), $\alpha$, is anomalously small $\left[ \alpha(T) \ll 10^{-5}\,{\rm K}^{-1} \right]$ over a wide range of temperature \cite{guillaume97,kussmann50}, a phenomenon known as the Invar effect. Their spontaneous volume magnetostriction, $w_{\rm s}$, measured at $T=0\,{\rm K}$ greatly exceeds that in body-centered-cubic (bcc) Fe and fcc Ni \cite{oomi81}. Their reduced magnetostriction, $w_{\rm s}/w_{\rm s}(0)$, scales with the square of the reduced magnetization, $\left[M/M(0)\right]^{2}$, up to a temperature near the Curie temperature, $T_{\rm C}$ \cite{oomi81,crangle63,sumiyama76,sumiyama79}. Surprisingly, only one of these two ferromagnets, namely Fe$_{0.65}$Ni$_{0.35}$, shows a peculiar thermal dependence of the reduced magnetization \cite{crangle63,sumiyama76}. 

Understanding all of the abovementioned phenomena within one framework is still a major open challenge. The most common theoretical explanation for the Invar effect involves the so-called 2$\gamma$-state model, where the iron atoms can switch between two magnetic states with different atomic volumes as the temperature is raised \cite{weiss63}. This theory, however, seems incompatible with the results of M\"ossbauer \cite{ullrich84} and neutron experiments \cite{brown01}. Another popular explanation emphasizes the importance of non-collinearity of the local magnetic moments on iron sites \cite{vanschilfgaarde99,dubrovinsky01}, though experiments undertaken to detect such non-collinearity have not found it \cite{cowlam03}. An alternative scenario with a purely magnetic origin for the Invar effect has been proposed \cite{kakehashi81}: the phenomenon is caused by anomalous thermal evolution of the magnitude of Fe moments. It is supported by a recent work on iron-platinum alloys \cite{khmelevskyi03} which involves {\it ab initio} density functional theory (DFT) calculations and the disordered local moment (DLM) model \cite{gyorffy85,johnson90}. However, the method employed in \cite{khmelevskyi03} cannot be extended to iron-nickel alloys. Thus, it is unable to provide a unified picture for the Invar effect in Fe$_{0.72}$Pt$_{0.28}$ and Fe$_{0.65}$Ni$_{0.35}$ and another treatment is called for.

A theoretical framework \cite{liot14} has recently been designed to address the spontaneous magnetization, the spontaneous volume magnetostriction, and their relationship in Fe$_{0.72}$Pt$_{0.28}$ and Fe$_{0.65}$Ni$_{0.35}$ in the temperature interval $0 \leq T/T_{\rm C} < 1$. Taking a similar approach as in \cite{khmelevskyi03} and \cite{liot12}, alloys in equilibrium at temperature $T$ have been modelled by random substitutional alloys in homogeneous ferromagnetic (FM) states, partially disordered local moment (PDLM) states, or DLM states depending on the fraction of Fe moments which are antiferromagnetically aligned with the magnetization at $T$, $x^{{\rm Fe}\downarrow}(T)$. The procedure could be divided into the following three stages. In the first stage, physical properties of interest (volume and magnetization) have been calculated for FM ($x^{{\rm Fe}\downarrow}=0$), PDLM ($0 < x^{{\rm Fe}\downarrow} < 1/2$), and DLM ($x^{{\rm Fe}\downarrow}=1/2$) states using {\it ab initio} DFT. In the second stage, the thermal evolution of the fraction of Fe moments which point down has been determined by noticing that an accurate description of the reduced magnetization is provided by a function of this form
\small 
\begin{eqnarray}\label{eqn1}
\frac{M(T)}{M(0)} = \bigg[ 1 -s \bigg( \frac{T}{T_{\rm C}} \bigg)^{3/2}- (1-s) \bigg( \frac{T}{T_{\rm C}} \bigg)^{p} \bigg]^{q}
\end{eqnarray}
\normalsize
and assuming that $x^{{\rm Fe}\downarrow}$ obeys the following equation
\small
\begin{eqnarray}\label{eqn2}
x^{{\rm Fe}\downarrow}(T) = \frac{1}{2} - \bigg[ \frac{1}{2} - x^{{\rm Fe}\downarrow}(0) \bigg] \bigg[ 1- \left( \frac{T}{T_{\rm C}} \right)^{p} \bigg]^{q}.
\end{eqnarray}
\normalsize
In the third and final step, the outputs from the previous steps have been combined to explore how the magnetization and the magnetostriction vary as the system is heated. Direct comparison between simulations results and experimental measurements has provided validation for the approach. The study supports the following ideas. The alloys at $T=0\,{\rm K}$ share several physical properties: the magnetization in a PDLM state collapses as the fraction of Fe moments which point down increases, following closely
\small
\begin{eqnarray}\label{eqn4}
M(0) - 2 M(0) x^{{\rm Fe}\downarrow},
\end{eqnarray}
\normalsize
while the volume shrinks, following closely 
\small
\begin{eqnarray}\label{eqn3}
V(0) - 4 [V(0)-V(1/2)] x^{{\rm Fe}\downarrow} (1-x^{{\rm Fe}\downarrow});
\end{eqnarray}
\normalsize
the volume in the FM state greatly exceeds that in the DLM state; $x^{{\rm Fe}\downarrow}(0)$ is close to 0. These common properties can account for a variety of intriguing phenomena displayed by both alloys, including the anomaly in the magnetostriction at $T=0\,{\rm K}$ and, more surprisingly perhaps, the scaling between the reduced magnetostriction and the reduced magnetization squared below the Curie temperature. However, the thermal evolution of the fraction of Fe moments which point down depends strongly on the alloy under consideration. This, in turn, can explain the observed marked difference in the temperature dependence of the reduced magnetization between the two alloys.

\begin{table}
\caption{\label{table1} The volume $V(0)$, the bulk modulus $B(0)$, and the Gr{\"u}neisen constant $\gamma(0)$ for Fe$_{0.72}$Pt$_{0.28}$, Fe$_{0.65}$Ni$_{0.35}$, and Fe$_{0.2}$Ni$_{0.8}$, according to EMTO calculations. All of these quantities are calculated for homogeneous ferromagnetic states.}
\begin{indented}
\item[]\begin{tabular}{@{}*{4}{c}}
\br
& volume (\AA$^{3}$) & bulk modulus (GPa) & Gr{\"u}neisen constant \\ 
\mr
Fe$_{0.72}$Pt$_{0.28}$ & 13.44 & 177 & 2 \\
Fe$_{0.65}$Ni$_{0.35}$ & 11.59 & 177 & 2 \\
Fe$_{0.2}$Ni$_{0.8}$ & 11.13 & 193 & 2 \\
\br
\end{tabular}
\end{indented}
\end{table}

This paper deals with the Invar effect in collinear ferromagnets Fe$_{1-x}A_{x}$ ($A={\rm Pt},{\rm Ni}$) with disordered fcc structure. The rich variety of thermal expansion displayed by these materials has firmly been established by experiments \cite{sumiyama79,tanji71}. This makes them particularly attractive for testing our general approach, identifying conditions under which an alloy shows the Invar effect, and investigating the mechanism of the Invar phenomenon. In principle, the LTEC can be derived from the configuration-averaged free energy which depends explicitly on volume and temperature. In practice, application of DFT to \emph{ab initio} calculations of a finite-temperature average free energy remains difficult, even in the adiabatic approximation where the electronic, the vibrational, and the magnetic contributions are treated separately. One of the major issues in implementing this strategy is how to incorporate magnetism correctly within the current approximations to the exchange and correlation functional \cite{abrikosov07}. Our simulation technique can be viewed as an extension of \cite{liot14} in which the vibrational contribution to the average free energy is treated within the Debye-Gr{\"u}neisen model \cite{moruzzi88,moruzzi90,herper99,crisan02}. Section~\ref{comp_methods} is devoted to computational details. Section~\ref{results} presents a comprehensive discussion of our results. As we shall see, this work challenges the conventional picture of the Invar effect as resulting from peculiar magnetic behaviour \cite{vanschilfgaarde99,dubrovinsky01,kakehashi81,khmelevskyi03,khmelevskyi05}. 

\begin{figure}
\includegraphics[width=8cm]{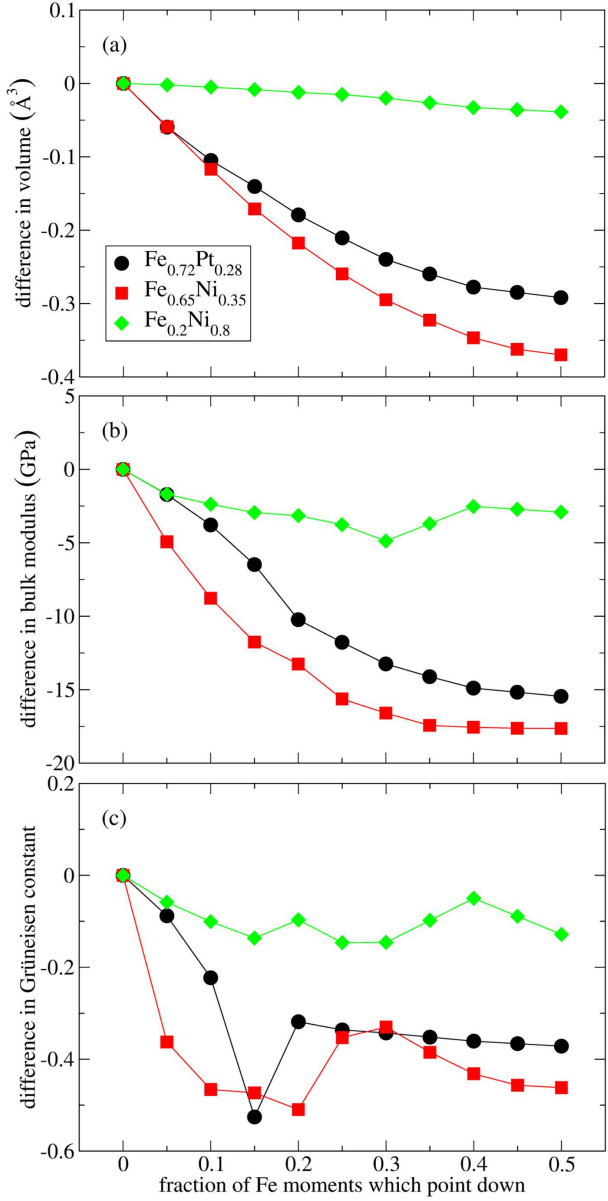}
\caption{The difference in volume $[V(x^{{\rm Fe}\downarrow})-V(0)]$ [panel~(a)], the difference in bulk modulus $[B(x^{{\rm Fe}\downarrow})-B(0)]$ [panel~(b)], and the difference in Gr{\"u}neisen constant $[\gamma(x^{{\rm Fe}\downarrow})-\gamma(0)]$ [panel (c)] plotted against the fraction of Fe moments which point down for Fe$_{0.72}$Pt$_{0.28}$, Fe$_{0.65}$Ni$_{0.35}$, and Fe$_{0.2}$Ni$_{0.8}$. Symbols show results of EMTO calculations. Note that the values for $V(0)$, $B(0)$, and $\gamma(0)$ are displayed in table~\ref{table1}.}
\label{figure1}
\end{figure}

\begin{figure}
\includegraphics[width=8cm]{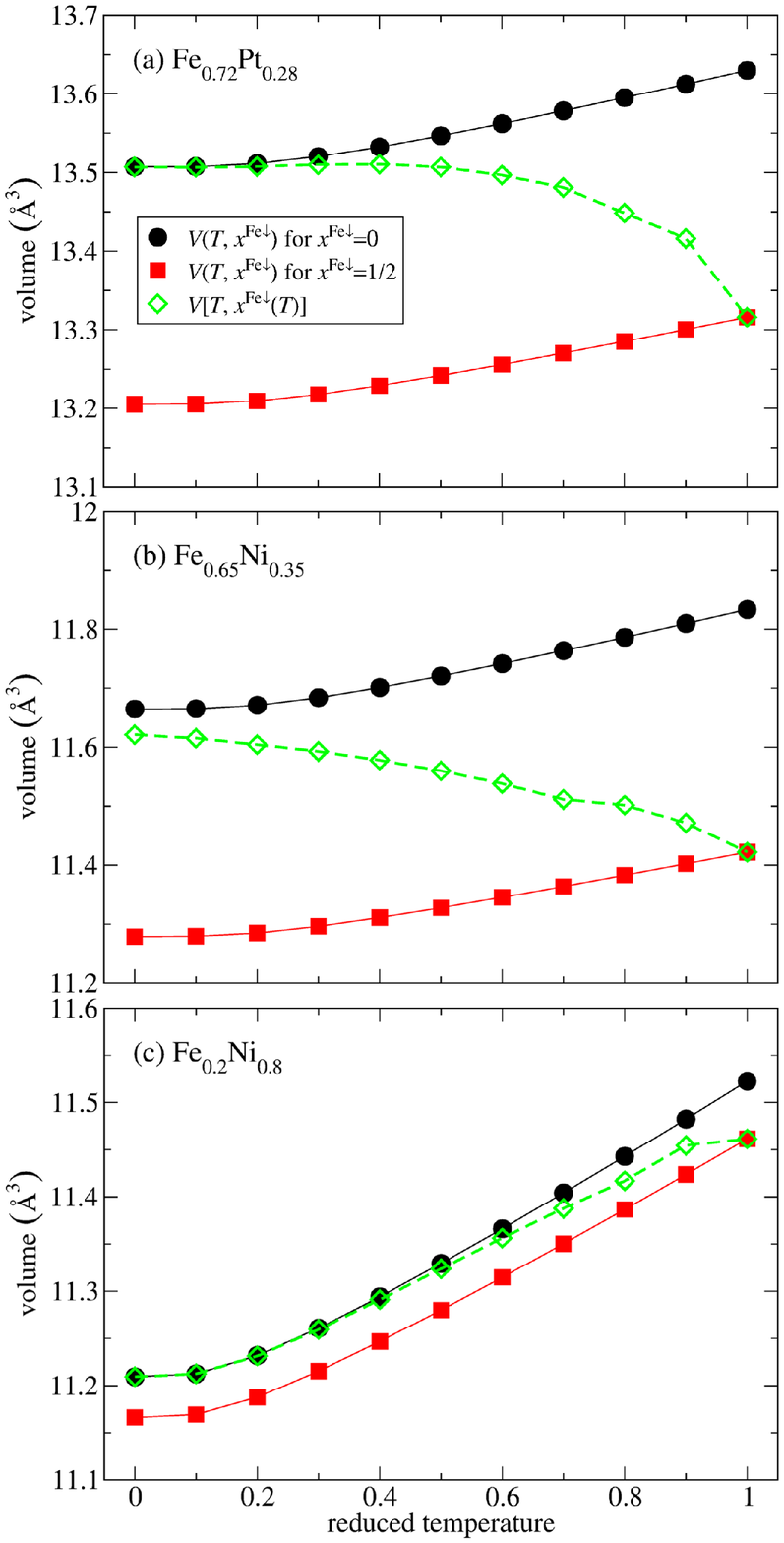}
\caption{The volumes $V(T,0)$, $V(T,1/2)$, and $V[T,x^{{\rm Fe}\downarrow}(T)]$ plotted against the reduced temperature $T/T_{\rm C}$ for Fe$_{0.72}$Pt$_{0.28}$ [panel (a)], Fe$_{0.65}$Ni$_{0.35}$ [panel (b)], and Fe$_{0.2}$Ni$_{0.8}$ [panel (c)].}
\label{figure2}
\end{figure}

\section{Computational methods}\label{comp_methods}

To address the Invar effect in collinear ferromagnets Fe$_{1-x}A_{x}$ ($A={\rm Pt},{\rm Ni}$) with disordered fcc structure, we extend the scheme developed in \cite{liot14} to include atomic vibrations. Fe$_{1-x}A_{x}$ alloys in equilibrium at temperature $T$ in the range $0 \leq T/T_{\rm C} < 1$ are modelled by random substitutional alloys in FM, PDLM, or DLM states depending on $x^{{\rm Fe}\downarrow}(T)$. The method remains divided into three main stages.

As a first step, we perform calculations of the volume $V(T, x^{{\rm Fe}\downarrow})$ for various temperatures and FM ($x^{{\rm Fe}\downarrow}=0$), PDLM ($0 < x^{{\rm Fe}\downarrow} < 1/2$), and DLM ($x^{{\rm Fe}\downarrow}=1/2$) states. For a fixed value of $T$ and $x^{{\rm Fe}\downarrow}$, the computational process is as follows:
\begin{enumerate}
\item We calculate the total energy $E ( r, x^{{\rm Fe}\downarrow} )$ for various Wigner-Seitz radii. This is done within the framework of the exact muffin-tin orbitals (EMTO) theory in combination with the full charge density (FCD) technique \cite{vitos01}. Further details can be found in \cite{liot14}. 
\item We deduce from the results of step (i) the Wigner-Seitz radius $r ( x^{{\rm Fe}\downarrow} )$, the volume $V(x^{{\rm Fe}\downarrow})$, the bulk modulus $B( x^{{\rm Fe}\downarrow} )$, and the Gr{\"u}neisen constant $\gamma ( x^{{\rm Fe}\downarrow} )$ \cite{moruzzi88}. 
\item For each Wigner-Seitz radius chosen in step (i), we estimate the contribution to the Helmholtz free energy $F_{\rm vib} ( T, r, x^{{\rm Fe}\downarrow} )$ from the outputs of step (ii)
\small
\begin{eqnarray}\label{eqn6}
F_{\rm vib}(T,r,x^{{\rm Fe}\downarrow}) = E_{\rm D} ( T, r, x^{{\rm Fe}\downarrow} ) - T S_{\rm D} ( T, r, x^{{\rm Fe}\downarrow} ),
\end{eqnarray}
\normalsize
where the vibrational energy and the vibrational entropy take the simple form
\small
\begin{eqnarray}\label{eqn7}
E_{\rm D} ( T, r, x^{{\rm Fe}\downarrow} ) = \frac{9}{8} k_{\rm B} \Theta ( r, x^{{\rm Fe}\downarrow} ) + 3 k_{\rm B} T D[ \Theta ( r, x^{{\rm Fe}\downarrow} )/T ]
\end{eqnarray}
\normalsize
and
\small
\begin{eqnarray}\label{eqn8}
S_{\rm D} ( T, r, x^{{\rm Fe}\downarrow} ) = 4 k_{\rm B} D[\Theta( r, x^{{\rm Fe}\downarrow})/T] - 3 k_{\rm B} \ln [ 1-e^{-\Theta( r, x^{{\rm Fe}\downarrow})/T}].
\end{eqnarray}
\normalsize
Here, $D$ denotes the Debye function. In analogy with \cite{moruzzi88,herper99}, we choose the Debye temperature $\Theta ( r,x^{{\rm Fe}\downarrow} )$ to be given by
\small
\begin{eqnarray}\label{eqn9}
\Theta ( r,x^{{\rm Fe}\downarrow} ) = \Theta_{0} ( x^{{\rm Fe}\downarrow} ) \bigg[ \frac{r( x^{{\rm Fe}\downarrow})}{r} \bigg]^{3 \gamma( x^{{\rm Fe}\downarrow})},
\end{eqnarray}
\normalsize
where $\Theta_{0} ( x^{{\rm Fe}\downarrow} )$ scales with $[r( x^{{\rm Fe}\downarrow}) B(x^{{\rm Fe}\downarrow})/M]^{1/2}$. We take the proportionality factor from \cite{herper99}.
\item We minimize the sum $E+F_{\rm vib}$ with respect to $r$ to obtain the volume $V(T, x^{{\rm Fe}\downarrow})$.
\end{enumerate}

As a second step, we investigate how heating the alloy affects its fraction of Fe moments which point down. The adopted method has already been described elsewhere \cite{liot14}.

In the third and final step, we combine the outputs from the two previous stages to explore how the volume $V [T,x^{{\rm Fe}\downarrow}(T)]$ and the anomalous contribution to the LTEC $\alpha_{\rm a}(T)$ vary as the temperature is raised. To allow for direct comparison between simulations and experiments \cite{wassermann90}, we conveniently define $\alpha_{\rm a}(T)$ as the difference between $\alpha(T)$ and $\alpha_{\rm n}(T)$, where the normal contribution to the LTEC measures the expansion that would occur if we heated the alloy in a DLM (`paramagnetic') state
\small
\begin{eqnarray}\label{eqn10}
\alpha_{\rm n}(T) = \Bigg[ \frac{1}{3V} \Bigg( \frac{\partial V}{\partial T} \Bigg)_{x^{{\rm Fe}\downarrow}} \Bigg] ( T, 1/2).
\end{eqnarray}
\normalsize
It is instructive to reexpress $\alpha_{\rm a}(T)$ as the sum of two terms
\small
\begin{eqnarray}\label{eqn11}
\alpha_{{\rm a},1}(T) = \Bigg[ \frac{1}{3 V} \Bigg( \frac{\partial V}{\partial T} \Bigg)_{x^{{\rm Fe}\downarrow}} \Bigg] [T, x^{{\rm Fe}\downarrow}(T)] - \Bigg[ \frac{1}{3 V} \Bigg( \frac{\partial V}{\partial T} \Bigg)_{x^{{\rm Fe}\downarrow}} \Bigg] (T, 1/2) 
\end{eqnarray}
\normalsize
and
\small
\begin{eqnarray}\label{eqn12}
\alpha_{{\rm a},2}(T) = \Bigg[ \frac{1}{3V} \Bigg( \frac{\partial V}{\partial x^{{\rm Fe}\downarrow}} \Bigg)_{T} \Bigg] [T, x^{{\rm Fe}\downarrow}(T)] \frac{dx^{{\rm Fe}\downarrow}}{dT}(T)
\end{eqnarray}
\normalsize
that corresponds to two distinct sources of anomaly: one associated with the expansion that would occur if we heated the alloy {\it without changing the configuration of Fe moments} and another one linked with the expansion that would occur if we changed the configuration of Fe moments, {\it but did not otherwise heat the system}. This latter contribution to $\alpha_{\rm a}(T)$ can be conveniently written as the product of the prefactor $-1/3$, the magnetostructural coupling
\small
\begin{eqnarray}\label{eqn13}
\kappa[T, x^{{\rm Fe}\downarrow}(T)] = \Bigg[ - \frac{1}{V} \Bigg( \frac{\partial V}{\partial x^{{\rm Fe}\downarrow}}\Bigg)_{T}\Bigg] [T, x^{{\rm Fe}\downarrow}(T)],
\end{eqnarray}
\normalsize
and the rate at which the fraction of Fe moments which point down fluctuates as the system is heated $dx^{{\rm Fe}\downarrow}/dT(T)$.  

\section{Results and discussion}\label{results}

According to experiments \cite{sumiyama79,tanji71}, Fe$_{0.72}$Pt$_{0.28}$, Fe$_{0.65}$Ni$_{0.35}$, and Fe$_{0.2}$Ni$_{0.8}$ exhibit a wide variety of thermal behaviour, the Fe-rich alloys showing the Invar effect and the Fe-poor alloy presenting thermal expansion similar to that of a paramagnetic compound. For this reason, they represent a suitable choice for testing the predictive power of the method developed in section~\ref{comp_methods}, formulating conditions for the occurrence of the Invar effect, and investigating the mechanism of the phenomenon. 

\subsection{Testing our approach}\label{resultsA}

\begin{figure}
\includegraphics[width=8cm]{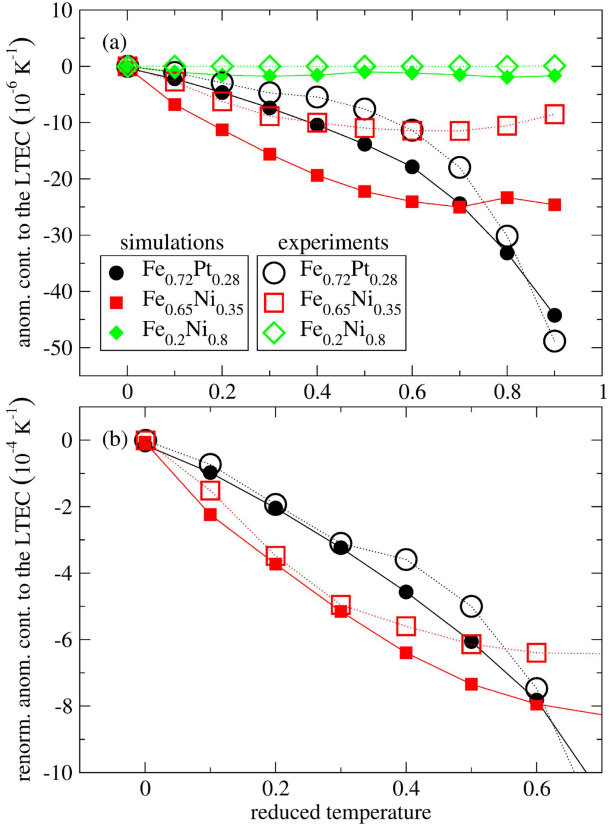}
\caption{Panel~(a): The anomalous contribution to the LTEC evaluated at temperature $T$ plotted against the reduced temperature for Fe$_{0.72}$Pt$_{0.28}$, Fe$_{0.65}$Ni$_{0.35}$, and Fe$_{0.2}$Ni$_{0.8}$. Panel~(b): The renormalized anomalous contribution for the two Fe-rich alloys. Filled symbols show results of numerical calculations. Open symbols display experimental data \cite{sumiyama79,tanji71,rellinghaus95,hayase73}.}
\label{figure3}
\end{figure}

Table~\ref{table1} shows the calculated volumes $V(0)$, bulk moduli $B(0)$, and Gr{\"u}neisen constants $\gamma(0)$. Figure~\ref{figure1} displays the calculated differences in volumes $[V( x^{{\rm Fe}\downarrow})-V(0)]$, bulk moduli $[B(x^{{\rm Fe}\downarrow})-B(0)]$, and Gr{\"u}neisen constants $[\gamma(x^{{\rm Fe}\downarrow})-\gamma(0)]$ for FM, PDLM, and DLM states. Note that the structural data have already been discussed \cite{liot14}. Regardless of the chemical nature of the alloy, the volume $V$ shrinks with increasing the fraction of Fe moments which point down, following closely (\ref{eqn3}). The volume for the FM state and the volume for the DLM state differ by more than $0.25\,{\rm \AA}^{3}$ in the Fe-rich alloys. The volume difference drops to $0.04\,{\rm \AA}^{3}$ when switching to the Fe-poor alloy. We now turn to describe the materials' response to uniform compression. Whether we consider Fe$_{0.72}$Pt$_{0.28}$, Fe$_{0.65}$Ni$_{0.35}$, or Fe$_{0.2}$Ni$_{0.8}$, the bulk modulus for the FM state lies within 175 and $195\,{\rm GPa}$. This is consistent with measurements performed on Fe$_{0.72}$Pt$_{0.28}$ and Ni \cite{oomi81-2}. The effect of raising $x^{{\rm Fe}\downarrow}$ on the bulk modulus $B$ mirrors to a certain extent that seen in panel~(a) for the volume $V$: (i) The bulk modulus decreases in the Invar alloys, revealing that these materials become easier to squeeze. (ii) The difference $[B(0)-B(1/2)]$, which amounts to $15$ in Fe$_{0.72}$Pt$_{0.28}$, $18$ in Fe$_{0.65}$Ni$_{0.35}$, and $3\,{\rm GPa}$ in Fe$_{0.2}$Ni$_{0.8}$, is considerably larger in the Fe-rich alloys. We note in passing that these findings might shed light on anomalies observed in measurements of bulk moduli \cite{dubrovinsky01,oomi81-2,manosa91,decremps04}. While we discuss figure~\ref{figure1}, we point out that numerical noise poses a significant problem for the determination of the Gr{\"u}neisen constants.

Figure~\ref{figure2} illustrates how the volumes $V(T,0)$, $V(T,1/2)$, and $V[T,x^{{\rm Fe}\downarrow}(T)]$ change with varying the temperature in the range $0 \leq T/T_{\rm C} < 1$. A useful way to analyze these data is as follows. Imagine that the magnetic configuration were fixed ($dx^{{\rm Fe}\downarrow}/dT=0$). Let us call the corresponding curve $V[T,x^{{\rm Fe}\downarrow}(0)]$; the curve for Fe$_{0.72}$Pt$_{0.28}$ and Fe$_{0.2}$Ni$_{0.8}$ is the uppermost black curve in panels~(a) and~(c). Then the material would not exhibit the Invar effect. This would also be the case if all of the curves $V(T,x^{{\rm Fe}\downarrow})$ for $0 \leq x^{{\rm Fe}\downarrow} \leq 1/2$ superimposed $[(\partial V / \partial x^{{\rm Fe}\downarrow})_{T}=0]$. In reality, however, raising the temperature from $T_{1}$ to $T_{2}$ causes the material to demagnetize, and the value of $x^{{\rm Fe}\downarrow}$ changes accordingly. One may say that the system hops from the curve $V[T,x^{{\rm Fe}\downarrow}(T_{1})]$ to the curve $V[T,x^{{\rm Fe}\downarrow}(T_{2})]$, resulting in a volume given by the curve $V [T,x^{{\rm Fe}\downarrow}(T)]$. This is shown as a dashed line. Insofar as panel~(b) allows us to judge for Fe$_{0.65}$Ni$_{0.35}$, each hop is to a curve lower than the last, cancelling the upward trend of each individual curve: this is the essence of the Invar effect. In section~\ref{resultsB}, we present a necessary condition under which an alloy shows the Invar effect. Consistent with the analysis of figure~\ref{figure2}, the criterion involves $\alpha_{{\rm a},2}=-1/3\,\kappa\,dx^{{\rm Fe}\downarrow}/dT$. 

In panel~(a) of figure~\ref{figure3}, we plot the calculated anomalous contribution to the LTEC $\alpha_{\rm a}(T)$ against the reduced temperature for Fe$_{0.72}$Pt$_{0.28}$, Fe$_{0.65}$Ni$_{0.35}$, and Fe$_{0.2}$Ni$_{0.8}$. Irrespective of the material under consideration, $\alpha_{\rm a}(T)$ exhibits a negative sign opposite to $\alpha_{\rm n}(T)$. However, only the Fe-rich materials possess the exceptional property that $\alpha_{\rm a}(T)$ compensates for $\alpha_{\rm n}(T)$ in a wide temperature range. Thus the approach predicts the occurrence of the Invar effect in Fe$_{0.72}$Pt$_{0.28}$ and Fe$_{0.65}$Ni$_{0.35}$ and its absence in Fe$_{0.2}$Ni$_{0.8}$. This perfectly matches experimental findings \cite{tanji71,rellinghaus95}. 

To further evaluate the predictive power of the method, we compare the calculated renormalized anomalous contribution to the LTEC $\tilde{\alpha}_{\rm a}(T) = \alpha_{\rm a}(T)/w_{\rm s}(0)$ with experimental observations \cite{sumiyama79,tanji71,rellinghaus95,hayase73} for the Invar alloys in panel~(b) of figure~\ref{figure3}. Note that we extract the calculated values for $w_{\rm s}(0) = \{ V[0,x^{{\rm Fe}\downarrow}(0)] - V(0,1/2) \} / V(0,1/2)$ from figure~\ref{figure2} and obtain 2.29\% for Fe$_{0.72}$Pt$_{0.28}$ and 3.03\% for Fe$_{0.65}$Ni$_{0.35}$. Panel~(b) of figure~\ref{figure3} reveals a good quantitative agreement between simulations and experiments. For instance, the curve for Fe$_{0.72}$Pt$_{0.28}$ intersects that for Fe$_{0.65}$Ni$_{0.35}$ at $T/T_{\rm C}=0.01$ and 0.6 according to simulations and $T/T_{\rm C}=0$ and 0.55 according to experiments. Another example involves the difference between $\tilde{\alpha}_{\rm a}(T)$ of the former alloy and that of the latter estimated at $T/T_{\rm C}=0.3$: The calculated quantity is $1.91\,10^{-4}\,{\rm K}^{-1}$, while the corresponding measured value amounts to $1.85\,10^{-4}\,{\rm K}^{-1}$.

Figure~\ref{figure3} provides strong evidence that the approach presented in this paper captures the essential physics of the Invar effect. This opens exciting opportunities for identifying conditions under which an alloy shows the Invar effect and investigating the mechanism of the phenomenon, which, in principle, can now be understood within the same framework as other intriguing observations \cite{liot14}, including: (i) the anomalously large magnetostriction in Fe$_{0.72}$Pt$_{0.28}$ and Fe$_{0.65}$Ni$_{0.35}$ at $T=0\,{\rm K}$, (ii) the peculiar temperature dependence of the reduced magnetization in Fe$_{0.65}$Ni$_{0.35}$, and (iii) the scaling of the reduced magnetostriction with the square of the reduced magnetization in Fe$_{0.72}$Pt$_{0.28}$ and Fe$_{0.65}$Ni$_{0.35}$ below the Curie temperature. 

\subsection{Identifying conditions under which an alloy shows the Invar effect}\label{resultsB}

\begin{figure}
\includegraphics[width=8cm]{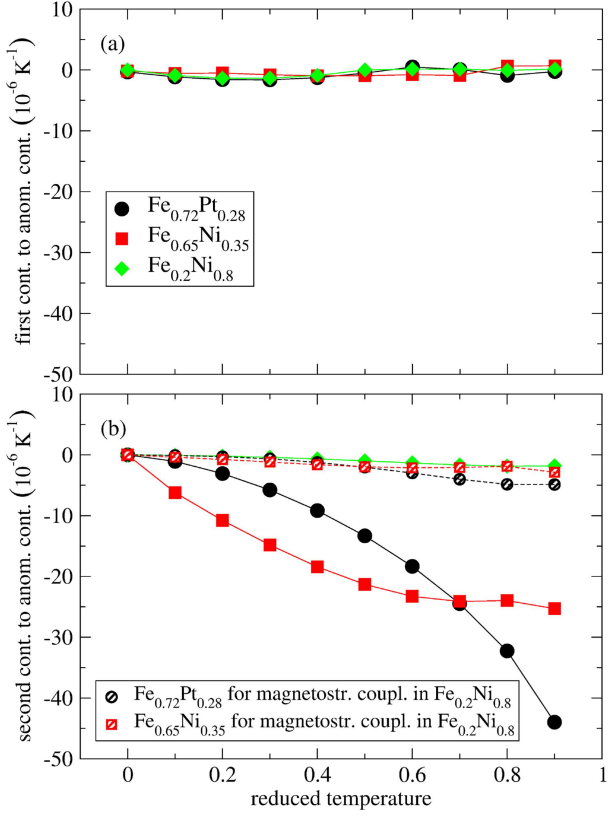}
\caption{The two contributions $\alpha_{{\rm a},1}(T)$ [panel~(a)] and $\alpha_{{\rm a},2}(T)$ [panel~(b)] to $\alpha_{\rm a}(T)$ plotted against the reduced temperature for Fe$_{0.72}$Pt$_{0.28}$, Fe$_{0.65}$Ni$_{0.35}$, and Fe$_{0.2}$Ni$_{0.8}$. Symbols show results of numerical calculations. Hatched symbols correspond to simulations performed for the two Fe-rich alloys with their magnetostructural coupling $\kappa$ substituted by that of Fe$_{0.2}$Ni$_{0.8}$.}
\label{figure4}
\end{figure}

The decomposition of the anomalous contribution to the LTEC $\alpha_{\rm a}(T)$ into its two parts $\alpha_{{\rm a},1}(T)$ and $\alpha_{{\rm a},2}(T)$ is plotted against $T/T_{\rm C}$ in figure~\ref{figure4} for Fe$_{0.72}$Pt$_{0.28}$, Fe$_{0.65}$Ni$_{0.35}$, and Fe$_{0.2}$Ni$_{0.8}$. The two competing terms $[ (1/3 V) (\partial V/\partial T)_{x^{{\rm Fe}\downarrow}}](T,1/2)$ and $[ (1/3 V) (\partial V/\partial T)_{x^{{\rm Fe}\downarrow}}][T, x^{{\rm Fe}\downarrow}(T)]$ balance each other almost completely, resulting in a very small $| \alpha_{{\rm a},1}(T) |$ (i.e., $| \alpha_{{\rm a},1}(T) |$ of the order of $10^{-6}\,{\rm K}^{-1}$, or less). It is clear that any strong deviation from zero shown by the anomalous contribution to the LTEC arises from $\alpha_{{\rm a},2}(T)=-1/3\,\kappa[T, x^{{\rm Fe}\downarrow}(T)]\,dx^{{\rm Fe}\downarrow}/dT(T)$. Features in the structural behaviour of the materials which have been observed experimentally (see figure~\ref{figure3}), but have remained unexplained, can now be interpreted on the basis of the abovementioned insight and our theoretical results displayed in figure~\ref{figure4}: (i) The drop in the anomalous contribution to the LTEC in Fe$_{1-x}$Ni$_x$ at $T/T_{\rm C}=1/2$ when the nickel concentration is reduced from 0.8 to 0.35 arises from the steep decrease of the product of the magnetostructural coupling $\kappa[T, x^{{\rm Fe}\downarrow}(T)]$ and the magnetic term $dx^{{\rm Fe}\downarrow}/dT(T)$. (ii) The fact that the anomalous contribution to the LTEC in Fe$_{0.72}$Pt$_{0.28}$ diminishes significantly as $T/T_{\rm C}$ is raised from 0.5 to 0.9, whereas that in Fe$_{0.65}$Ni$_{0.35}$ does not reflects the different behaviours of $\kappa\,dx^{{\rm Fe}\downarrow}/dT$ in this interval: this physical quantity decreases drastically in the Fe-Pt case, but remains almost constant in that of Fe-Ni.

On the basis of figures~\ref{figure3} and~\ref{figure4}, we argue that {\it the Invar phenomenon occurs only when the thermal expansion arising from the temperature dependence of the fraction of Fe moments which point down $\alpha_{{\rm a},2}$ compensates for the thermal expansion associated with the anharmonicity of lattice vibrations $\alpha_{\rm n}$ in a wide temperature interval}.

\begin{figure}
\includegraphics[width=8cm]{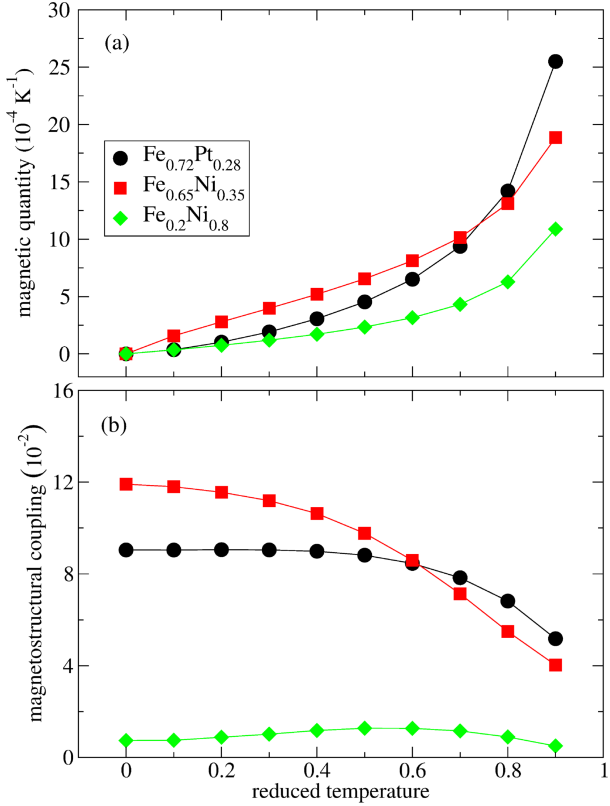}
\caption{The magnetic quantity $dx^{{\rm Fe}\downarrow}/dT(T)$ [panel~(a)] and the magnetostructural coupling $\kappa[T, x^{{\rm Fe}\downarrow}(T)]$ [panel~(b)] plotted against the reduced temperature for Fe$_{0.72}$Pt$_{0.28}$, Fe$_{0.65}$Ni$_{0.35}$, and Fe$_{0.2}$Ni$_{0.8}$.}
\label{figure5}
\end{figure}

A natural question to ask is: Why do some alloys fulfill this necessary condition for the occurrence of the Invar effect and others do not? To shed light on this matter, consider our results presented in figures~\ref{figure4} and~\ref{figure5}. In Fe$_{0.2}$Ni$_{0.8}$, the magnetostructural coupling is weak at $T=0\,{\rm K}$ ($\kappa[0, x^{{\rm Fe}\downarrow}(0)]=0.74\,10^{-2}$) and $\alpha_{{\rm a},2}$ fails to counterbalance $\alpha_{\rm n}$ over a broad temperature range. In the Fe-rich alloys, however, the magnetostructural coupling is especially strong ($\kappa[0, x^{{\rm Fe}\downarrow}(0)] > 9\,10^{-2}$) and $\alpha_{{\rm a},2}$ compensates for $\alpha_{\rm n}$ in a wide temperature interval. Interestingly, if we substitute their magnetostructural coupling $\kappa$ by that of Fe$_{0.2}$Ni$_{0.8}$, the physical situation changes drastically, resembling that in Fe$_{0.2}$Ni$_{0.8}$. This supports the idea that only alloys with strong magnetostructural coupling at $T=0\,{\rm K}$ can show the Invar effect. 

\section{Conclusion} \label{conclusion}

To address the Invar effect in collinear ferromagnets Fe$_{1-x}A_{x}$ ($A={\rm Pt},{\rm Ni}$) with disordered fcc structure, we have extended the scheme developed in \cite{liot14} to include atomic vibrations. Fe$_{1-x}A_{x}$ alloys in equilibrium at temperature $T$ in the range $0 \leq T/T_{\rm C} < 1$ have been modelled by random substitutional alloys in FM, PDLM, or DLM states depending on $x^{{\rm Fe}\downarrow}(T)$. The method has been divided into three main stages. As a first step, we have performed calculations of the volume $V(T, x^{{\rm Fe}\downarrow})$ for various temperatures and FM, PDLM, and DLM states. As a second step, we have investigated how heating the alloy affects its fraction of Fe moments which point down. In the third and final step, we have combined the outputs from the two previous stages to explore how the volume $V [T,x^{{\rm Fe}\downarrow}(T)]$ and the anomalous contribution to the LTEC $\alpha_{\rm a}(T)$ vary as the temperature is raised. It is worth emphasizing that neither partial chemical order \cite{crisan02} nor static ionic displacement \cite{liot06,liot09,liot_thesis} has been explicitly taken into account at any stage.

Tests results for Fe$_{0.72}$Pt$_{0.28}$, Fe$_{0.65}$Ni$_{0.35}$, and Fe$_{0.2}$Ni$_{0.8}$ have provided evidence that the methodology captures the essential physics of the Invar effect. This opens exciting opportunities for investigating the mechanism of the phenomenon, which, in principle, can now be understood within the same framework as other intriguing observations \cite{liot14}. 

We have decomposed the anomalous contribution to the LTEC $\alpha_{\rm a}$ into two parts and studied each of them separately, for Fe$_{0.72}$Pt$_{0.28}$, Fe$_{0.65}$Ni$_{0.35}$, and Fe$_{0.2}$Ni$_{0.8}$. Our results support the following criterion: The Invar phenomenon occurs only when the thermal expansion arising from the temperature dependence of the fraction of Fe moments which point down $\alpha_{{\rm a},2}$ compensates for the thermal expansion associated with the anharmonicity of lattice vibrations $\alpha_{\rm n}$ in a wide temperature interval. 

Finally, based on the study of $\alpha_{{\rm a},2}$ and $\kappa$, we have predicted that only alloys with strong magnetostructural coupling at $T=0\,{\rm K}$ can show the Invar effect. This work challenges the conventional picture of the Invar effect as resulting from peculiar magnetic behaviour.

\ack
The author thanks I. A. Abrikosov (Link{\"o}ping), B. Alling (Link{\"o}ping), C. A. Hooley (St Andrews, U.K.), A. E. Kissavos (Link{\"o}ping), and J. Neugebauer (D{\"u}sseldorf) for fruitful discussions.


\end{document}